\documentclass[reprint, amsmath, amssymb, aip, jcp, longbibliography]{revtex4-2}

\usepackage[english]{babel}
\usepackage[dvips]{graphics}
\usepackage{graphicx,epsfig}
\usepackage{amsmath}
\usepackage{amsfonts}
\usepackage{amssymb}
\usepackage{xcolor}
\usepackage{multirow}
\usepackage[normalem]{ulem}
\usepackage{booktabs}
\usepackage{lineno}
\usepackage{url}
\usepackage{breakurl}
\usepackage{subcaption}
\usepackage{placeins}  
\usepackage{amsthm}

\usepackage{xcolor}

\usepackage[breaklinks,
            colorlinks=true,
            citecolor=blue,
            linkcolor=blue,
            urlcolor=blue]{hyperref}

\usepackage[justification=raggedright,singlelinecheck=false]{caption}

\newcommand{\LiCN}{Li-CN}
\newcommand{\LiNC}{Li-NC}

\newcommand{\CR}[1]{\textcolor{red}{#1}}

\newcommand{\Tcn}{\mathcal{T}_{\rm kin}^{\rm CN}(0)}

\makeatletter
\AtBeginDocument{%
  \def\@bibdataout@aps{}%
  \def\@bibdataout@aip{}%
}

\begin{document}


\title{%
Transition states and dynamical bottlenecks in the\\
\texorpdfstring{{\LiNC}\(\rightleftharpoons\){\LiCN}}{{\LiNC} <-> {\LiCN}}
isomerizing reaction with the Lagrangian betweenness
}

\author{P. Garc\'ia-Cuadrillero}
\affiliation{Grupo de Sistemas Complejos, Departamento de Matem\'atica Aplicada,
   Escuela T\'ecnica Superior de Edificaci\'on,
   Universidad Polit\'ecnica de Madrid,
   Avenida Juan de Herrera 6, 28040 Madrid, Spain.}

\author{F. Revuelta}
\affiliation{Grupo de Sistemas Complejos,
   Escuela T\'ecnica Superior de Ingenier\'ia Agron\'omica,
   Alimentaria y de Biosistemas,
   Universidad Polit\'ecnica de Madrid,
   Avenida Puerta de Hierro 2, 28040 Madrid, Spain.}

\author{R. M. Benito}
\affiliation{Grupo de Sistemas Complejos,
   Escuela T\'ecnica Superior de Ingenier\'ia Agron\'omica,
   Alimentaria y de Biosistemas,
   Universidad Polit\'ecnica de Madrid,
   Avenida Puerta de Hierro 2, 28040 Madrid, Spain.}

\author{F. Borondo}
\affiliation{Departamento de Qu\'imica,
   Universidad Aut\'onoma de Madrid,
   Cantoblanco, 28049 Madrid, Spain.}

\date{\today}

\begin{abstract}
We study the vibrational dynamics that is responsible 
for {\LiNC}\(\leftrightharpoons\){\LiCN} isomerization reaction using 
the Lagrangian betweenness,
a tool recently developed to identify dynamical bottlenecks in the context of oceanographic transport.
Following the spirit of the
betweenness-centrality metric,
which efficiently identifies the nodes that act
as bottlenecks in complex networks,
the Lagrangian betweenness is similarly capable to identify
bottlenecks in dynamical systems.
On the one hand,
in this paper we show
the ability of the Lagrangian betweenness to
identify transition states,
which are short-lived states that are formed
at the top of the energetic barrier that the system must surpass
in order to transition
the reaction  from
the (initial) reactants to the (final) products.
On the other hand,
we also demonstrate that the Lagrangian betweenness is also
able to identify other dynamical bottlenecks,
and phase-space invariant structures such as 
manifolds,
chains of islands and invariant tori,
which are also responsible for the 
intricate
vibrational dynamics
of the system.
\end{abstract}

\maketitle

%
\section{Introduction}
\label{sec.intro}

The deeper understanding of
molecular reactions has intrigued generations of chemists throughout history. Achieving the capacity to steer chemical transformations
-to favor particular reaction channels~\cite{Lique15, Falcinelli20, Xu25},
to suppress undesired side reactions~\cite{Schmidt06, May20},
or to modulate rates by external intervention~\cite{Park20}-
has long been a central aspiration in chemistry.
Over the past few decades,
the advent of tools such as shaped ultrafast laser pulses~\cite{Korolkov97, Tseng11, Townsend21},
tailored electromagnetic fields~\cite{Lemeshko13} and cavities~\cite{Li21},
molecular beam methods~\cite{Lee87, vandeMeerakker12},
and advanced computational control strategies~\cite{Baltussen24, Sharma24}
has created new opportunities
to manipulate the transformation of initial reactants into desired products
in a more selective and efficient manner.

Regardless of the system complexity,
most chemical transformations share a fundamental dynamical feature:
the transient formation of a metastable configuration known
as the transition state (TS) or activated complex~\cite{Marcelin15, Eyring31, Evans38}.
A remarkable exception are roaming reactions~\cite{Townsend04, Krajnak18}, 
where part of the reactants (or fragments of them) wander 
flat regions of the  potential energy surface (PES)~\cite{Marcelin15, Eyring31}
rendering to the products,
without crossing any energetic barrier.
In general,
however,
this barrier unfailingly separates the reactant and product basins,
and it must be surmounted to proceed with the reaction.
Then, 
the TS can be subsequently formed
close to the saddle point that is located at its top,
or at the top of the free energy surface.
The TS constitutes a fleeting configuration
that represents, furthermore,
the most important 
bottleneck for the reaction to take place.

Being the rate-limiting step,
the key ingredients of reactivity are encoded by the TS.
Its first direct observation dates back to 
a groundbreaking experiment performed 
by Lovejoy \emph{et al.} in 1992~\cite{Lovejoy92, Marcus92},
where the vibrational levels of the TS
that is formed during ketene dissociation were 
observed \emph{in situ} for the first time.
However,
the precise characterization of TS remains
still nowadays far from trivial~\cite{Chen19}.
The challenge relies on the low living times
of the TSs,
which typically also present
a very complex and intriguing dynamics.

In addition to small molecules,
 the ability to identify and control TSs
is also critical in the realm of biomolecular conformational dynamics,
like allostery or protein folding.
However,
in high-dimensional PESs
or free-energy landscapes typical of proteins or complex molecular systems,
identifying the true TS is extraordinarily difficult.
In these cases,
the barrier regions are sparsely sampled,
the topology is intricate,
and naive dimensionality reduction or clustering may fail to
capture the relevant 
bottleneck regions where the potential TSs can be formed.
Still, 
recent studies have sought to detect TS
conformations in proteins from molecular dynamics data using machine learning frameworks, treating TS conformations as out-of-distribution states relative to metastable basins~\cite{Liu2025},
and to predict TS structures in materials using data‐driven classification methods~\cite{Kim2025}.

Since the TS is the critical bottleneck for reactions to take place~\cite{Haenggi90, Pechukas81, Truhlar96, Miller98},
it is not surprising that,
ever since its inception~\cite{ Eyring38a, Eyring38b, Evans38, Wigner38, Wigner39},
TS theory (TST) has mainly focused on its study.
The theory has not only succeeded in providing 
the rate of rupture of chemical bonds~\cite{Evans01} 
in single molecules,
but also in providing rates in
much larger and complex systems, 
such as proteins~\cite{Chung12, Schuler13}
and RNA~\cite{Dudko06} unfolding,
and the crossing rates of
DNA through entropic barriers~\cite{Vestergaard16}.

In summary,
TST 
assumes
the existence of a 
surface of no return (dividing surface)
that separates reactants from product and it is crossed
once and only once by all reactive trajectories.
Then,
the reaction rate is given by
the ratio between reactive flux through the surface
and the reactive population,
which also depends on the barrier height,
as Arrhenius foresaw.
However,
the identification of a perfect recrossing-free surface is
far away from trivial,
and remains as an active field of research.
Consequently,
the existence of recrossings cause the TST rate to
overestimate the true rate,
sometimes strongly.
This recrossing problem leads to the development of a variational 
version of TST~\cite{Truhlar80, Bao17},
whose primary goal consisted in minimizing the
number of recrossings.

From a dynamical perspective,
the bottleneck character of the TS
is rather ubiquitous,
as Wigner remarked~\cite{Wigner38}.
As a consequence,
the TST perspective
is also suitable to unravel the 
\emph{reactivity} in any other system whose
phase-space can be partitioned in two different
regions separated by an energetic barrier.
That is the paradigmatic case of the (chaotic) ionization of Hydrogen atom 
in the presence of external pulses~\cite{Jaffe99, Uzer02, Schweiner15, Dong17}.
Saddle points, and then TSs,
do also play a central role in the dynamics and
rearrangement of clusters formed by several atoms~\cite{Hinde93},
and in the
spin dynamics driven by magnetic fields
\cite{Maihoefer22, Moegerle22}.
Remarkably,
TST has been also successfully applied to study
the diffusion jumps in solids~\cite{Toller85},
and
conductance due to ballistic electron transport through microjunctions~\cite{Eckhardt95}.
Within much larger scales,
TS-like behavior is observed
in the \emph{escaping dynamics}~\cite{Wiggins92chaotic}
that describes the
ship capsize in rough seas~\cite{Naik17}.

Within a classical description,
in isolated systems  with two degrees of freedom
(DoF),
the TS corresponds to a periodic orbit (PO) located at
the top of the potential energy barrier.
Moreover,
this TS defines a non-recrossing DS in configuration space
known as PODS,
as recognized by Pechukas, MacLafee, and Pollak, 
\cite{Pechukas73, Pollak78, Pechukas79}
which can be used to compute exact reaction rates.
When projected onto configuration space,
the PODS connects two branches of the equipotentials.
Likewise,
the invariant manifolds of this PODS partition the energy shell
in phase space,
this effectively 
separating the regions that enable reaction from those that do not.
The definition of a suitable set of coordinates allows the
construction of PODS even under
velocity-dependent forces, like Coriolis forces~\cite{Jaffe99, Jaffe00}.

In higher dimension,
the TS generalizes to a normally hyperbolic invariant manifold (NHIM),
which acts as a \emph{mutidimensional} saddle point
at the top of the energetic barrier~\cite{Wiggins01, Uzer02, Waalkens04}.
The invariant manifolds that emerge from the NHIM organize the phase-space transport,
mediating the flow of trajectories between reactants and products
(or between successive intermediates).

In this work, 
we show the ability of the Lagrangian betweenness (LB)
to identify bottlenecks in
the {\LiCN}\(\leftrightharpoons\){\LiNC} isomerization reaction.
The LB is a novel tool recently proposed by
Ser-Giacomi \emph{et al.} 
to identify 
bottlenecks in dynamical systems.
Inspired by the networks-theory betweenness-centrality metric,
the authors demonstrated
in the Ref.~\onlinecite{SG21}
the ability of the LB to 
characterize hidden circulation regimes in
realistic geophysical ﬂows.
%
Thus,
due to the \emph{bottleneck} behavior of the TSs
in chemical systems,
it is natural to test the performance of the LB in
a realistic molecular system.
We show that 
LB
successfully identifies the main 
TSs  that mediate the isomerization,
as well as secondary bottlenecks that arise due to formation of
partial dynamical barriers in phase space.
Moreover,
we also show that the LB is also capable of describing other phase-space
structures which play a central role in the system dynamics.
The results rendered by the LB are compared with other
well-established chaos indicators,
such as the Lagrangian descriptors (LDs)~\cite{Madrid09, Lopesino15}
and the finite-time Lyapunov exponents (FTLEs)~\cite{Haller01, Haller02},
showing perfect agreement,
but also remarking the better performance of LB
to unveil the system bottlenecks.
Then,
this work can be regarded as a proof of principle that demonstrates that
the LB is an efficient
and interesting tool that might be of interest to identify
TSs in more complex molecular or even biological systems
with higher dimensional PESs.

Furthermore,
{\LiNC}$\rightleftharpoons${\LiCN}
isomerization
considered
provides an excellent benchmark to test the ability of 
the LB since it is described by 
an \emph{ab-initio} PES
and 
presents a small number of DoF,
providing a tractable yet rich platform to explicitly visualize phase-space structures and unravel fundamental dynamical mechanisms in a realistic molecular system.
Consequently,
the classical and quantum phase space of 
{\LiNC}$\rightleftharpoons${\LiCN}
has been extensively characterized.
On the one hand,
it is a paradigm for classical mixed phase-space dynamics,
a situation where
regular islands coexist  at the same energy with chaotic regions,
and has been studied through non-linear dynamics~\cite{Losada08, Benitez13, Benitez15, Revuelta19, Revuelta21}
and multifractal analysis \cite{Tarquis01}.
On the other hand,
this mixed phase space leads to subtle quantum-classical correspondence phenomena, including the quantum manifestation of the Poincaré-Birkhoff theorem and scarring phenomena \cite{Arranz96, Arranz98, Arranz10, Wisniacki11}.
Likewise,
beyond isolated-molecule dynamics, the system has proven invaluable for studying chemical reactivity in condensed phases. In particular, embedding the {\LiNC}$\rightleftharpoons${\LiCN} isomerization in a thermal bath has enabled direct numerical observation of the celebrated Kramers turnover in reaction rates \cite{GM08, GM12}. Furthermore, the presence of saddle-point resonances in this bound, chaotic system offers key insights into energy transport and transition-state dynamics \cite{GomezLlorente92}.

The remaining of this article is organized as follows.
First,
we introduce in Sec.~\ref{sec:system}
two Hamiltonian models for the system under study.
Second,
we briefly review the conceptual background of 
the LB  and its mathematical formulation in
Sec.~\ref{sec:method}.
Third,
we validate the LB by comparison with other phase-space indicators
in Sec.~\ref{sec:results},
which contains the main findings and the corresponding discussion.
Finally, we summarize and conclude in Sec.~\ref{sec:conclusions}.

\section{System description}
\label{sec:system}

\begin{figure}[t]
    \centering
    \includegraphics[width=0.9\columnwidth]{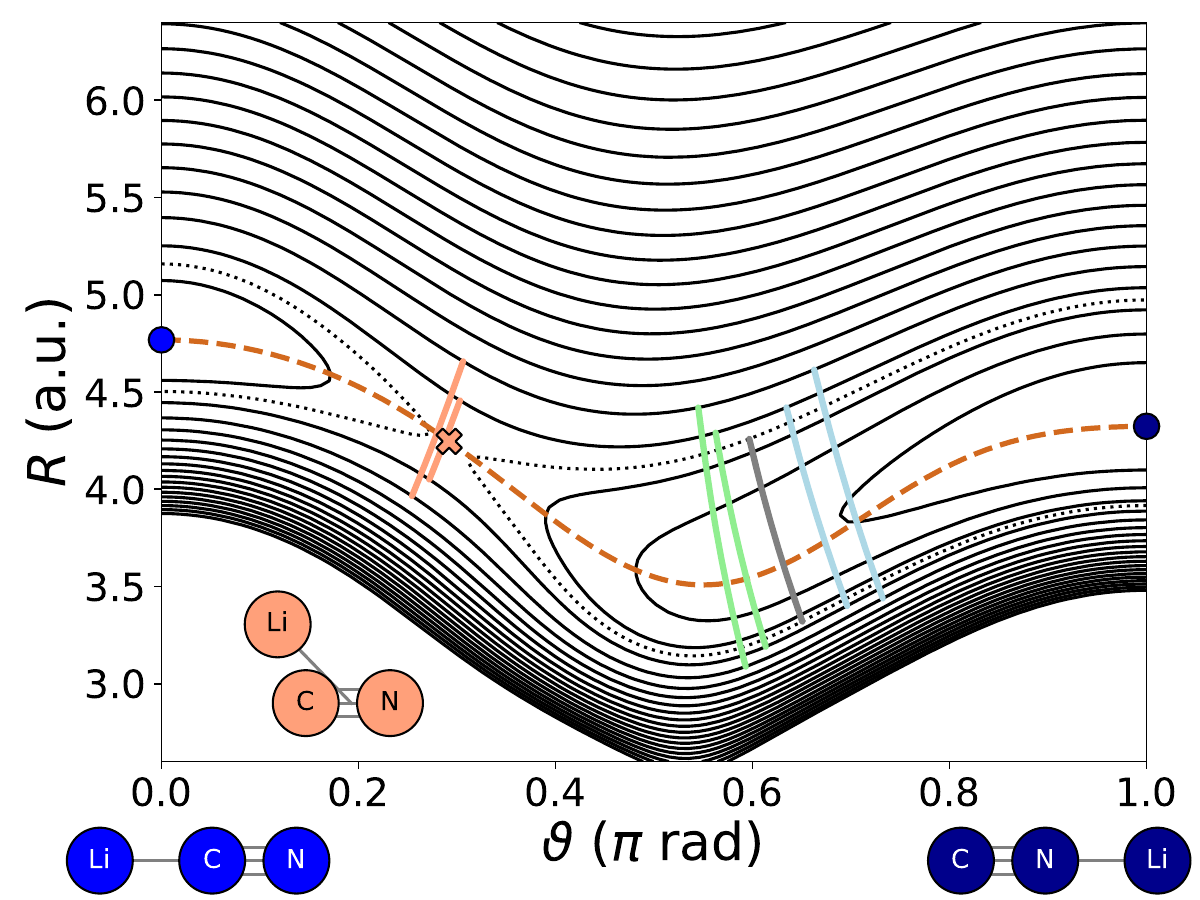}
    \caption{
Contour plot
of the \emph{ab-initio} potential energy surface
$V_{\mathrm{Li\text{-}CN}}(R,\theta)$ of the {\LiCN} molecular system, 
shown at $1000\ \mathrm{cm^{-1}}$ intervals (solid black),
and
for
$E_{\mathrm{bif}} = 3440.6\ \mathrm{cm^{-1}}$ (dotted).
The minimum-energy path (dashed brown),
seven characteristic periodic orbits (solid almost vertical lines),
the saddle-point 
(light orange cross),
the two potential minima
(blue circles),
along with
sketches of the corresponding molecular configurations
are also superimposed.
}
    \label{fig:pes}
\end{figure}

The system under study
corresponds to the isomerization reaction  
{\LiNC}\,\(\leftrightharpoons\)\,{\LiCN}, considered in its rotational ground state (\(J = 0\)).  
In this regime, 
as sketched in 
Fig.~\ref{fig:pes},
the vibrational dynamics can be fully 
characterized using the Jacobi coordinates
$\{R, \theta, r\}$,
where \(R\) denotes the Li--CN stretching coordinate, \(\vartheta\) is the Li--CN bending angle, and \(r\) represents the C$\equiv$N stretching coordinate.
The corresponding Hamiltonian~\cite{Borondo95b}
 is given by
\begin{equation} \label{eq:H3dof}
\mathcal{H}_3 = \frac{P_R^2}{2\mu_1} + \frac{P_\vartheta^2}{2\,I_\vartheta(R,r)} + \frac{P_r^2}{2\mu_2} +
V(R,\vartheta,r),
\end{equation}
where~$P_R$, $P_\vartheta$,
and~$P_r$
are the conjugate momenta,
$\mu_1 = m_{\text{Li}} (m_{\text{C}} + m_{\text{N}})/ (m_{\text{Li}} + m_{\text{C}} + m_{\text{N}})$,
and
$\mu_2 = m_{\text{C}}\,m_{\text{N}}/(m_{\text{C}} + m_{\text{N}})$,
are the reduced masses,
which depend on the masses of the atoms of
the molecule,
and
$I_\vartheta(R,r) = \left[1/(\mu_1 R^2) + 1/(\mu_2 r^2)\right]^{-1}$
the moment of inertia associated with the bending motion
of the Li atom around the center of mass of the C-N group.
Finally,
the last term in Eq.~\eqref{eq:H3dof} is
the PES
defined as
\begin{equation}
V(R,\vartheta,r) = V_{\text{Li-CN}}(R,\vartheta) + V_{\text{CN}}(r),
\end{equation}
where \(V_{\text{Li--CN}}(R,\vartheta)\) describes the interaction between the Li atom
and the CN dimmer
shown in Fig.~\ref{fig:pes}.
The C-N bound
is modeled 
with the following Morse potential
\begin{equation}
V_{\text{CN}}(r) = D\left[1 - e^{-\beta(r - r_e)}\right]^2,
\end{equation}
with~$D = 0.29135\,\text{a.u.}$,
$\beta = 1.4988\,\text{a.u.}$,
and~$r_e = 2.186\,\text{a.u.}$
the C-N bond equilibrium distance.

As sketched in Fig.~\ref{fig:pes},
the potential
$V_{\text{Li--CN}}(R,\vartheta)$ presents two
well minima,
where the molecule is structured
in one of the stable collinear isomers
{\LiCN}
for~$\vartheta = 0$\,rad
(left blue circle),
or 
{\LiNC}
for~$\vartheta = \pi$\,rad
(right blue circle),
being the latter more stable due to its smaller energy
(0 vs 2281\,cm$^{-1}$).
These two stable configurations,
which are also sketched at the bottom of the figure,
are  separated by an energetic barrier
which has a saddle point
located at~$(\vartheta, R) =
(0.292\pi\,\textnormal{rad}, 4.22\,\textnormal{a.u.})$
(light orange cross),
being its energy 3455\,cm$^{-1}$.
As the energy increases,
the saddle point becomes a PO that
progressively shifts to the left.
This behavior is evident from the
comparison of the
two light orange POs
located in the vicinity of the saddle point,
which have
vibrational energies of
$4000\,\mathrm{cm^{-1}}$
and
$5000\,\mathrm{cm^{-1}}$, 
respectively,
as inferred from the positions of their turning points,
where~$(P_\vartheta, P_R)=(0,0)$.

We note that
due to the triple bond of the C-N dimmer,
the~$r$ coordinate does not change much,
while the changes in the~$R$ coordinate
are much more substantial due to the
loose behavior in this DoF.
Then, 
one can effectively set~$r \equiv r_e$,
so that~$P_r=0$,
and then the 3-DoF Hamiltonian given by
Eq.~\eqref{eq:H3dof} can be reduced to
\begin{equation} \label{eq:H2dof}
\mathcal{H}_2 = \frac{P_R^2}{2\mu_1} + \frac{P_\vartheta^2}{2\,I_\vartheta(R,r_e)} + 
V(R,\vartheta).
\end{equation}

To conclude,
notice in Fig.~\ref{fig:pes}
the minimum energy path (MEP)
 \(R = R_e(\vartheta)\) 
 that joins the two stable
minima 
through the saddle point.


\section{Tools and methods}
\label{sec:method}

This section is divided in two parts.
First,
we summarize in Sec.~\ref{sec:chaos_indicators}
the chaos indicators used 
for validation purposes of the LB results.
Second,
we briefly review
in Sec.~\ref{sec:LB}
the main characteristics of the
LB  originally introduced in 
the Ref.~\onlinecite{SG21},
which is subsequently used to 
identify the energetic and dynamical bottlenecks
that are formed in the
{\LiNC}$\rightleftharpoons${\LiCN}
isomerization reaction.
%
%



\subsection{Description of the
chaos indicators considered}
\label{sec:chaos_indicators}

Here,
we present the three chaos indicators considered 
to test the validity of the LB results:
the Poincar\'e surface of section (PSoS)
reported in Sec.~\ref{sec:PSoS},
the LDs
discussed
in Sec.~\ref{sec:LD}, 
and the FTLEs
summarized in Sec.~\ref{sec:FTLE}.


\subsubsection{The Poincar\'e surface of section}
\label{sec:PSoS}

The PSoS
distinguishes among one of the most popular chaos indicators
developed~\cite{Arnold06, LL10, Celletti10, Cvitanovic16}.
This tool unambiguously distinguishes regular from irregular or chaotic motion
in 2-DoF systems, 
and provides valuable information on 
systems with 3 DoF.
In summary,
chaotic trajectories are represented as an infinite set of
disconnected points on the Poincaré map
as their motion is only restricted to be confined
within the corresponding energy shell.
Contrarily,
periodic and quasiperiodic orbits 
are shown very differently.
On the one hand,
POs
are marked  as a finite set of points on the PSoS.
On the other hand,
quasiperiodic orbits are represented on the PSoS
as finite sets of closed curves that surround 
fixed points associated with POs.
These lines correspond to \emph{cuts} or \emph{sections}
of the invariant tori which further restrict the motion.

Among all possibilities,
a PSoS defined along the MEP of Fig.~\ref{fig:pes}
offers the optimal one for the case study
(see Ref.~\onlinecite{Arranz10} for further details).
However,
in order to make the map
area-preserving one must also perform the following canonical transformation
\begin{subequations}
\begin{align}
\rho &= R - R_e(\vartheta), \label{eq:can1}\\[1mm]
P_\rho &= P_R, \label{eq:can2}\\[1mm]
\psi &= \vartheta, \label{eq:can3}\\[1mm]
P_\psi &= P_\vartheta - \left(\frac{dR_e}{d\vartheta}\right) P_R. \label{eq:can4}
\end{align}
\label{eq:PSoS}
\end{subequations}
The PSoS~\eqref{eq:PSoS}
is defined as the set of points in the \((\psi, P_\psi)\) plane for which the new coordinate \(\rho=0\) and the time derivative \(\dot{\psi} > 0\).
%
The PSoS is very efficient in identifying invariant tori which
restrict regular motion
in 2-DoF systems.
However,
this tool is not capable to identify other phase-space structures
which lie within the chaotic regions of phase-space.
Thus,
in order to validate the LB results reported in Sec.~\ref{sec:results},
we have also used the two complementary tools introduced below:
the LDs and the FTLEs.
Both of them were
initially developed to study
 fluid and geophysical flows~\cite{Madrid09, Haller01},
but
their current applications have also extended to other fields,
such as
open maps~\cite{Carlo20},
celestial mechanics~\cite{CG21, Flores26}, 
or
chemistry~\cite{Craven15, Craven16, Patra18,
Revuelta19, Krajniak20, Revuelta21, Revuelta23}.

\subsubsection{The Lagrangian Descriptors}
\label{sec:LD}
LDs were first introduced by 
Jim\'enez Madrid and Mancho to characterize
the phase-space structure of fluid flows~\cite{Madrid09}.
In their original formulation,
the LD was associated with the arc length
of a trajectory over a finite integration time.
This initial definition was later refined by 
considering $p$ norms.
For a system with
$N$ equations of motion (EoM),
i.\,e.,
$N/2$ DoF,
this new expression is given by~\cite{Lopesino15}
\begin{equation} \label{eq:LD}
    M(\mathbf{z_0}, \tau) = \sum_{i=0}^N \int_0^\tau \vert \dot z_i (t) \vert^p dt,
     \quad \textnormal{with}~0 < p \le 1,
\end{equation}
where~$\dot{z}_i$ is 
the ~$i^\textnormal{th}$ EoM.
For Hamiltonian systems, like the molecular system considered in this work
[cf. Eqs.~\eqref{eq:H3dof} and~\eqref{eq:H2dof}],
the EoM follow Hamilton’s equations
\begin{equation} \label{eq:EoM}
    \dot z_i = \mathbf{J} \frac{\partial \mathcal{H}}{\partial z_i},
     \quad \textnormal{with}~\mathbf{J}=\left(
    \begin{array}{cc}
           \mathbf{0}_{N/2 \times N/2} & \mathbf{I}_{N/2 \times N/2}  \\
         - \mathbf{I}_{N/2 \times N/2} & \mathbf{0}_{N/2 \times N/2} 
    \end{array}
    \right) .
\end{equation}

In both the original and the $p$‑norm formulations, it was soon observed that,
for sufficiently large integration times~$\tau$,
LDs exhibit strong differences between initial 
conditions lying on invariant manifolds 
and those on neighboring points.
Thus,
comparing their values over a sufficiently dense grid of initial conditions
(ICs),
the  stable and unstable 
manifolds can be straightforwardly identified
\cite{Madrid09, Mendoza10, Mancho13, Lopesino15}.

\subsubsection{The finite-time Lyapunov exponents}
\label{sec:FTLE}

Given a flow map
$\Phi^t_{t_0}(\mathbf{z_0}) = \mathbf{z} (t; t_0, \mathbf{z_0})$
generated by the corresponding EoM
[cf. Ec.~\eqref{eq:EoM}],
the linearized evolution of an infinitesimal perturbation is given by the tangent map $D\Phi^\tau(x_0)$.
The FTLE~\cite{Haller01, Haller02} associated with
an IC
$\mathbf{z_0}$ and time $\tau$ is given by
\begin{equation} \label{eq:FTLE}
    \lambda(\mathbf{z_0},\tau)=\frac{1}{ \vert \tau \vert  }\,\ln\!\big(\sqrt{\sigma_{\max} (C) )}\big),
\end{equation}
with~$\sigma_{\max} (C)$
the largest eigenvalue of the
Cauchy-Green strain tensor
$C = \left( \nabla \Phi^t_{t_0}(\mathbf{z_0}) \right)^T  \nabla \Phi^t_{t_0}(\mathbf{z_0}) $,
which measures the maximal stretching factor experienced
by any unit perturbation over the time interval~$\tau$.
Positive values of $\lambda(\mathbf{z_0},\tau)$ indicate
finite-time exponential separation of trajectories,
while values near zero reflect regular or weakly unstable behavior.
As in the case of the LDs,
FTLEs fields (the spatial distribution of $\lambda$ over
a sufficiently dense set of ICs)
are widely used to reveal transport barriers and phase-space structures.


\subsection{The Lagrangian betweenness}
\label{sec:LB}
Ever since its inception,
the betweenness-centrality metric
has
played a central role
in network theory~\cite{Freeman77,Newman09}.
A node with high betweenness centrality lies on many of the
most important routes linking other nodes and
therefore strongly influences the overall connectivity
of the network.
This result can be a consequence of
(i) the existence of a bottleneck due to the privilege position of the node,
which, e.g., is the only one that links two regions of the network,
or
(ii) the high number of connections of the node with the rest of the network.
In directed networks, 
the connections between nodes have a specific orientation.
Thus,
the ability of the node~$i$ to be
reached by the other nodes 
within a time interval~$\tau$,
is quantified by its in-degree
$K_i^I(0, \tau)$,
whereas its ability to reach the others is given by its
out-degree $K_i^O(0, \tau)$.
Then,
the likelihood of connecting two nodes
between a given one
in such a way that the node~$i$ is crossed at time~$t \in [0, \tau]$
is given by 
$K_i^I(0, t) K_i^O(t, \tau-t)$.
As a consequence,
the probability of
the node~$i$ to be crossed
within the time interval~$[0, \tau]$
can be estimated as
%
\begin{equation}
\label{eq:intKiKo}
\frac 1\tau \int_0^\tau K_i^I(0, t) K_i^O(t, \tau-t) dt .
\end{equation}
The previous reasoning has been recently extended to continuous dynamical systems
by modeling it with a
sufficiently high number of nodes~\cite{SG21}.
In this case, 
the rate of attraction/repulsion to each IC,
i.\,e., node,
is determined by its corresponding invariant manifolds.
On the one hand,
trajectories will be
attracted to that point of phase space in the direction(s) of the stable manifold(s),
but they will be simultaneously separated from it
in the direction(s) of the unstable manifold(s)
at a rate that can be estimated with the FTLEs
(see further details in Sec.~\ref{sec:chaos_indicators}).
Then,
the in- and out-degrees
can be associated with 
\begin{eqnarray}
K_i^I(0, \tau) &\approx& e^{ \tau\lambda(\mathbf{z}_i; \tau,-\tau) } , \label{eq:Ki} \\
K_i^O(0, \tau) &\approx& e^{ \tau\lambda(\mathbf{z}_i; 0,\tau) } , \label{eq:Ko}
\end{eqnarray}
respectively,
with 
\(\lambda\bigl(\mathbf{z}_i; t_0,\tau\bigr)\)
is the FTLE
for the IC
\(\mathbf{z}_i\),
starting at time $t_0$ and
evaluated over a time interval $\tau$.
Finally,
by substitution of Eqs.~\eqref{eq:Ki} and~\eqref{eq:Ko}
in Eq.~\eqref{eq:intKiKo},
we finally arrive at the definition of the LB
provided by
\begin{equation}
\label{eq:LB}
B_i^{L}(0, \tau) \;=\; \frac 1\tau
\int_{0}^{ \tau} 
e^{ t\,\lambda(\mathbf{z}_i; t,-t)}
e^{   ( \tau - t)\,\lambda(\mathbf{z}_i; t, \tau-t) } \,dt.
%
\end{equation}

Equation~\eqref{eq:LB} was originally introduced within the field of
oceanic transport~\cite{SG21},
where it could be successfully used to identify the dynamical bottlenecks
in the Adriatic sea and in the Kerguelen region of the Indic ocean.
In the next section,
we will show the ability of the LB to also identify the
reactive bottlenecks of {\LiCN}$\rightleftharpoons${\LiCN} isomerization
described in Sec.~\ref{sec:system},
extending the range of validity
of this interesting tool to a realistic chemical reaction.

\section{Results and discussion}
\label{sec:results}
This section presents the main findings of our study
of the {\LiNC}$\rightleftharpoons${\LiCN} isomerizing reaction.
It is divided in two parts.
First,
Sec.~\ref{sec:results_2d}
reports on the results obtained for the 2-DoF model.
Second,
Sec.~\ref{sec:results_3d} addresses
the investigation for the 3-DoF model.

\subsection{Model with two degrees of freedom}
\label{sec:results_2d}

This section is devoted to
 the
2-DoF system 
 described by the
Hamiltonian~\eqref{eq:H2dof}. 
We employ the LB
with the ICs~$(\psi, P_\psi)$
on the PSoS,
which correspond to a four-dimensional vectors
~$\mathbf{z}_0 = (R_0, \theta_0, P_{R,0}, P_{\theta,0})$,
whose components can be calculated 
with the help of Eq.~\eqref{eq:PSoS}
once the vibrational energy is fixed.
In Sec.~\ref{sec:results_2d_ts},
we study
 the  TS
located
at the top of the energetic barrier separating
the two stable isomers.
Next,
in Sec.~\ref{sec:results_2d_sn}, we examine
the saddle-node bifurcation that is responsible for a dynamical
barrier that 
effectively hinders the isomerization process~\cite{Borondo95, Borondo96, Zembekov97, Revuelta21}.

\subsubsection{Transition state revealed by Lagrangian betweenness}
\label{sec:results_2d_ts}

\begin{figure}[!ht]
    \centering
    \includegraphics[width=0.9\columnwidth]{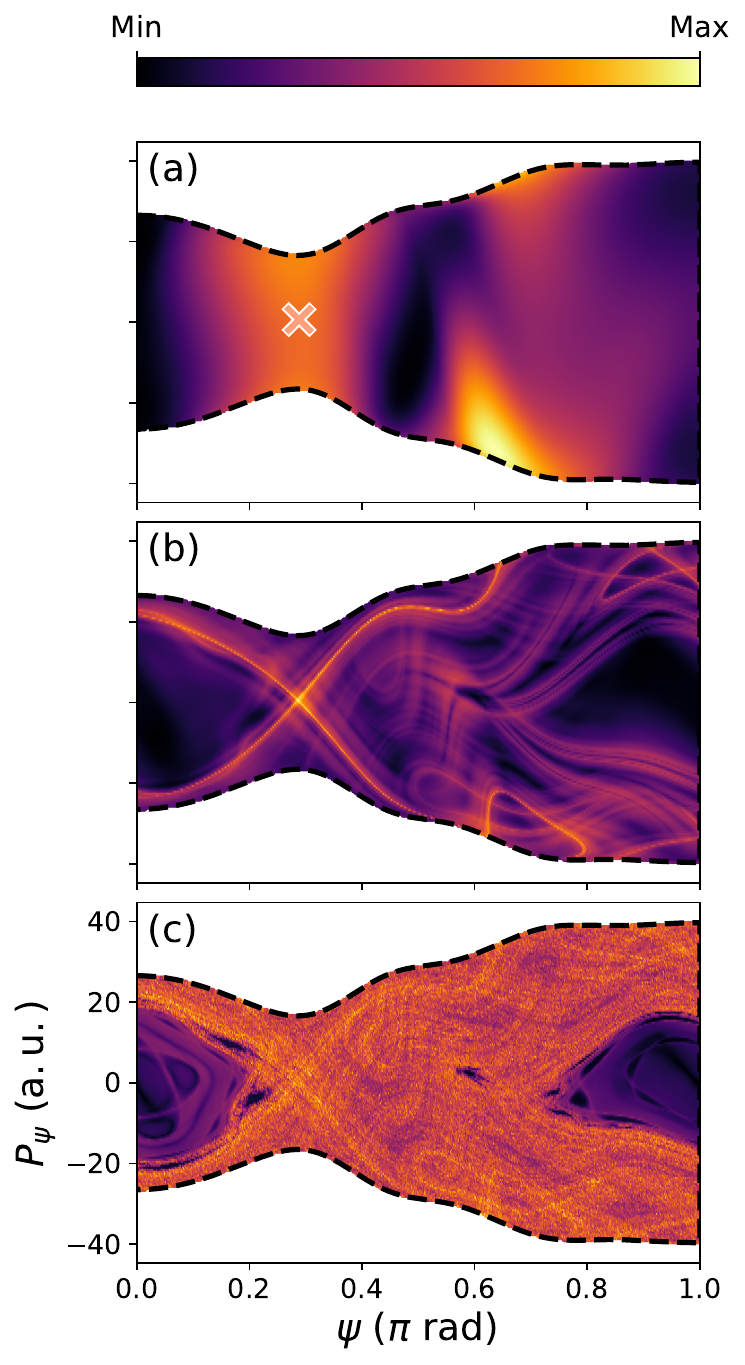}
    \caption{%
    Value of the Lagrangian betweenness~\eqref{eq:LB}
   (in logarithmic scale)
    on the Poincaré surface of section~\eqref{eq:PSoS} 
    for the  2-degrees-of-freedom Hamiltonian~\eqref{eq:H2dof}
    with a vibrational energy of $E = 4000\,\mathrm{cm}^{-1}$
    and an integration time of 
    (a)~\(\tau = 10^3\) a.u.,
    (b)~\(\tau = 10^4\) a.u.,
    and
    (c)~\(\tau = 10^5\) a.u.
    The black dashed line corresponds to
    the energy boundary.
    The light orange cross in panel (a)
    marks the position of the hyperbolic point associated with the transition state of
    Fig.~\ref{fig:pes}
    (short light orange line).
}
    \label{fig:LB_2d_e4000}
\end{figure}

To begin with,
we examine the effect of the integration time~$\tau$ on the LB
given by Eq.~\eqref{eq:LB}.
For this goal,
we show in
Fig.~\ref{fig:LB_2d_e4000} 
the LB
for a uniform set of ICs
on the
PSoS given by Eq.~\eqref{eq:PSoS}
with~$\rho=0$ and $\dot \psi>0$
for a vibrational energy of \(4000\,\mathrm{cm}^{-1}\) and three different
computational times.
As can be seen,
if the integration time is too low,
like in the case of~$\tau=10^3$\,a.u. shown in
Fig.~\ref{fig:LB_2d_e4000}(a),
the LB is given by a slow varying function, 
where no structure can be recognized.
However, 
if the integration time is sufficiently increased,
the LB successfully identifies the phase-space bottlenecks
that are formed due to
the underlying phase-space structures.
Indeed,
$\tau = 10^4$\,a.u. is a suitable time,
as inferred from visual inspection of
Fig.~\ref{fig:LB_2d_e4000}(b).
Notice the clear ``$\times$'' structure that is 
visible in the vicinity of the saddle point that is located
at the top of the energetic barrier~$\psi = 0.292 \pi$\,rad,
which separates
the two {\LiCN} and {\LiNC} isomers
(see Fig.~\ref{fig:pes}).
The crossing point of 
the mentioned structure is the hyperbolic point
shown in Fig.~\ref{fig:LB_2d_e4000}(a),
which corresponds to the
shortest
light orange PO that is shown in Fig.~\ref{fig:pes}.
As the system has 2 DoF,
this PO corresponds to the TS,
and effectively acts as a totally recrossing-free
dividing surface~\cite{Pechukas73, Pollak78, Pechukas79}.
The lines that emerge from the hyperbolic point
are the invariant manifolds of the PO,
which fully explain the reaction mechanism.
Namely,
in the close vicinity of the TS,
the ICs
approximate (separate)
in the direction of the stable (unstable) manifold~\cite{Cvitanovic16}.

Let us stress two important points.
First,
notice the logarithmic scale in Fig.~\ref{fig:LB_2d_e4000}, 
which implies that the values of the LB along the invariant manifolds of the
TS [cf. Fig.~\ref{fig:LB_2d_e4000}(b)] are several orders of magnitude larger than for other ICs
considered on the PSoS.
Further apart from the TS,
the manifolds fold and unfold creating a very intricate structure,
the homoclinic tangle,
that is responsible for the highly complex dynamics.
Remarkably,
the manifolds on other regions of the PSoS can be 
also identified,
but the value of the LB is much smaller than in the vicinity of the TS.

Second,
though the
adequate integration time of
$\tau = 10^4$\,a.u. = 2481.89\,fs
considered in Fig.~\ref{fig:LB_2d_e4000}(b)
might look at first glance too large, it
is only five times larger than
the period of the TS
(49.23 fs),
which is the shortest PO of the system.
As in the case of other chaos integrators, 
a suitable integration time~$\tau$
must be larger than the inverse of the stability exponent~$\lambda_u$
of the PO associated with the TS, which equals
$1/\lambda_u = 885$\,a.u., i.e.,
it is 12 times smaller than the considered integration time.
(see further details in Ref.~\onlinecite{Revuelta21}).
When considering smaller integration times,
the underlying phase-space structures
that determine the
system behavior do not have enough time 
to yield such a noticeable imprint on the dynamics.
Conversely,
if this time is too large,
the local information of the manifolds is incorporated.
Figure~\ref{fig:LB_2d_e4000}(c) shows an example.
In this case, 
the LB has been computed for
$\tau = 10^5$\,a.u.,
which unravels the invariant structures 
in other regions of the phase space. 
At first glance,
the invariant manifolds associated with the TS
that are remarkably visible in
Fig.~\ref{fig:LB_2d_e4000}(b)
are not observed in this case.
A closer look, however,
shows that the 
some of the points of the ``$\times$'' structure
have very large values of the LB,
but the manifolds are not so easily visible because they
have very different values along them.
However,
the invariant tori within the phase-space regions
where the motion is mostly regular
are now clearly identified because of their different LB values
compared to the ones along the invariant manifolds that embrace them.
Notice in Fig.~\ref{fig:LB_2d_e4000}(c)
the presence of a  large torus around the stable isomer {\LiCN} located at
$(\psi, P_\psi)=(0, 0)$\,(rad,\,a.u.).
At the center of this torus,
a stable stretching PO corresponding to a 1:1 resonance is found. 
For such a $n_R:n_\vartheta$ resonance,
the value in the $R$ coordinate changes $n_R$ times
while that in the $\vartheta$ coordinate changes $n_\vartheta$ times
after one period of time.
The torus encloses an infinite hierarchy of progressively smaller tori,
arranged in an onion-like configuration.
At their common center,
the resonance is manifested as an elliptic fixed point on the PSoS
located at coordinates $(0, 0)$\,(rad,\,a.u.).
The motion of ICs
on the tori is quasiperiodic,
and mimics that of the 1:1 resonance.
Namely,
after a time interval of the 1:1 resonance,
the $\theta$ and $R$ coordinates reach
values close to the initial ones.

As can be seen,
another invariant torus, 
manifested as two stable islands in Fig.~\ref{fig:LB_2d_e4000}(c)
surrounds the previous one.
These stable islands
surround a 1:2 resonance,
which is associated with a periodic motion where the molecule changes two times in
the~$\theta$ coordinate before changing once in the~$\theta$ coordinate.
Another chain of regular islands can be also identified within the {\LiCN} well
much closer to the boundary of regular motion
(external part of the cyan region),
which is related to a 1:5 resonance.
Similarly,
the position of the main stable isomer 
{\LiNC} located at
$(\psi, P_\psi)=(\pi, 0)$\,(rad,\,a.u.), coincides with a stable
stretching 1:1 PO.
Two additional chains of islands
corresponding to 1:4 and 1:3 resonances
are located between
the 1:1 and 1:5 resonances.
Infinite series of invariant tori are present in between,
which prescribe
quasiperiodic motion.

\begin{figure}[t]
    \centering
    \includegraphics[width=0.95\columnwidth]{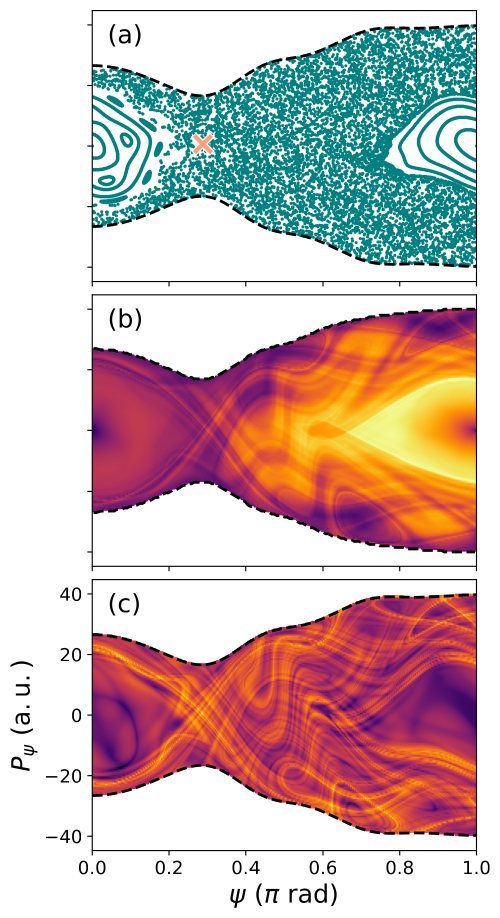}
    \caption{%
     Results
    for the 2-degrees-of-freedom Hamiltonian~\eqref{eq:H2dof}
    with \(E = 4000\,\mathrm{cm}^{-1}\)
    yielded by
    the Poincaré surface of section~(a),
    and the sums of the forward and backward
    time evolutions up to \(\tau = 10^4\) a.u.
    of the Lagrangian descriptors (b),
    and
    of the finite-time Lyapunov exponent (c).
    The light orange cross in panel (a)
    marks the position of the hyperbolic point associated with the transition state
}
    \label{fig:FTLE_FLI_SALI_LD}
\end{figure}

To demonstrate the precision and suitability of the LB
to identify chemical TSs,
and other phase-space structures,
we show in Fig.~\ref{fig:FTLE_FLI_SALI_LD} the results for the other three
well-established chaos indicators:
the composite PSoS (a),
and
the sums of the forward and backward 
FTLEs (b) [cf.~Eq.\eqref{eq:FTLE}]
and
LDs (c) [cf.~Eq.\eqref{eq:LD}].
To guarantee a proper comparison,
the same vibrational energy
and,
in the case of 
three indicators,
a computational time  of
Fig.~\ref{fig:LB_2d_e4000}
were considered, i.\,e.,
\(E=4000\,\mathrm{cm}^{-1}\) and \(\tau=10^4\) a.u.

First,
notice
that most of the PSoS
of Figs.~\ref{fig:FTLE_FLI_SALI_LD}(a)
is  filled by a dense sea of disconnected points,
this indicating a strong prevalence of chaotic motion in most 
the phase space~\cite{LL10, Celletti10, Cvitanovic16}.
The PSoS is not able to identify the invariant manifolds of the TS,
whose position is highlighted as a light orange cross,
nor the ones of any other PO of the system
within that region.
Conversely,
the PSoS adequately identifies the invariant tori that determine
regular motion close to the two stable isomers of the system
located at the potential minima.
The comparison between
Figs.~\ref{fig:LB_2d_e4000}(c)
and~\ref{fig:FTLE_FLI_SALI_LD}(a)
shows that the LB is also able to identify these structures
when computed for sufficiently long times.
The reason lies on the different dynamical behavior
present for the trajectories that move over invariant tori,
and those  that surround them,
whose dynamics is determined by the invariant manifolds that emerge from the 
unstable fixed points that lie between them
as the Poincaré-Birkhoff theorem prescribes~\cite{LL10}.

Second,
the accurate performance of the LB
in the identification of
the reactive bottleneck, \emph{aka} TS,
is assessed by comparison of
Fig.~\ref{fig:LB_2d_e4000}(b)
with
Figs.~\ref{fig:FTLE_FLI_SALI_LD}(b)
and~\ref{fig:FTLE_FLI_SALI_LD}(c),
which
show the
results for the FTLE and for the LD, respectively.
As can be seen,
the the hyperbolic structure associated with the TS
is present in all figures.
Notice, however, that the FTLE and the LDs present
much complex landscapes.
Indeed,
both of these tools reveal 
not only the manifolds close to the TS but also in very distant regions
of phase space,
which make the identification of the TS more involved than with the LB.

\begin{figure}[t]
    \centering
     \includegraphics[width=0.9\columnwidth]{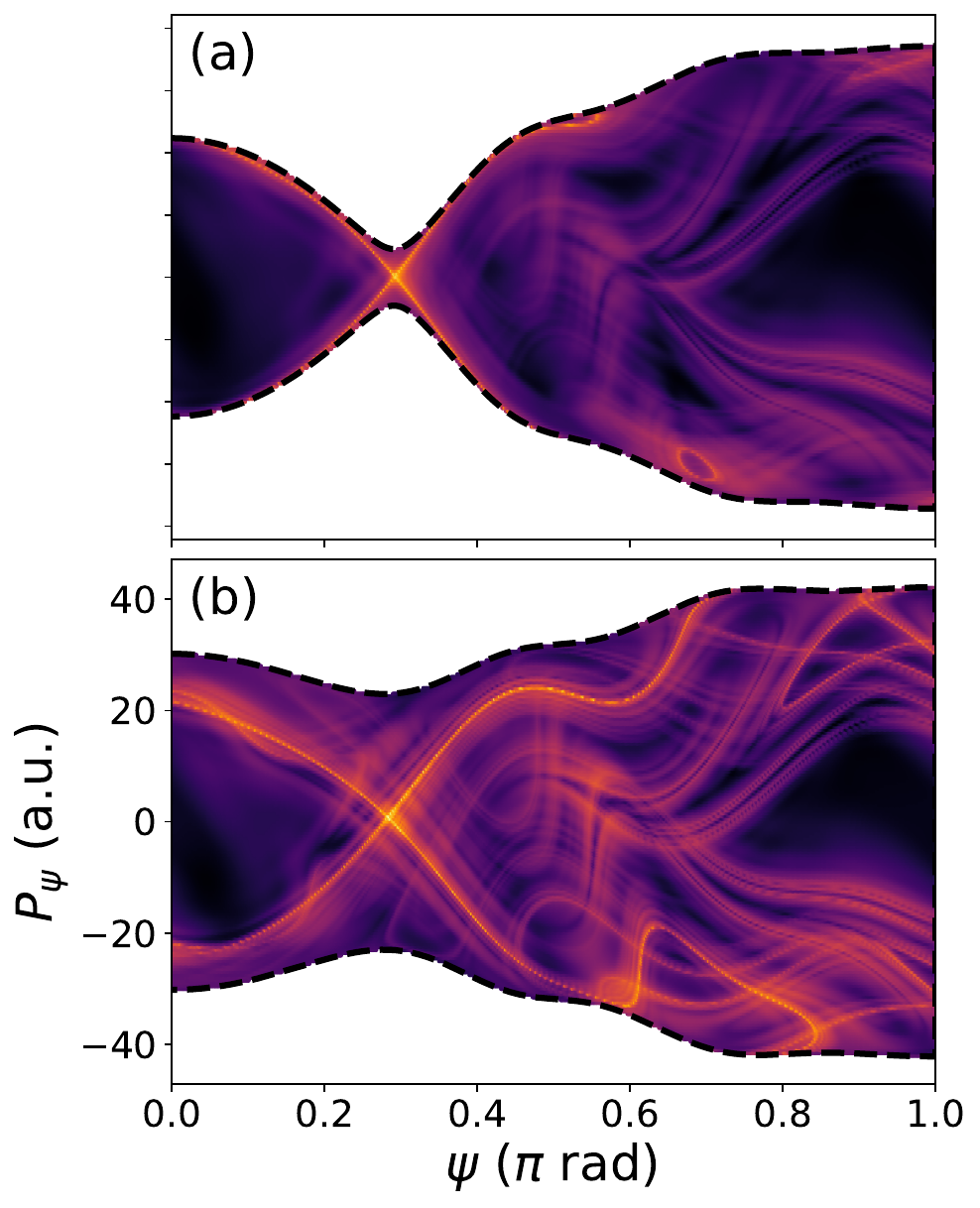}
    \caption{%
    Same as Fig.~\ref{fig:LB_2d_e4000}
    for \(\tau = 10^4\) a.u., and
    (a)~\(E=3500\,\mathrm{cm}^{-1}\), and
    (b)~\(E=4500\,\mathrm{cm}^{-1}\).
}
    \label{fig:LB_2d_varyE}
\end{figure}

Next,
in order to study the ability of the LB
to identify TSs at other vibrational energies,
we show in Fig.~\ref{fig:LB_2d_varyE} 
the LB field for two different
vibrational energies,
smaller and larger than the one considered so far,
and for the same computational time
of \(\tau = 10^4\)\,a.u.
As expected,
the accessible phase space strongly depends
on the vibrational energy.
On the one hand,
for
\(3500\)\,\(\mathrm{cm}^{-1}\) 
[see Fig.~\ref{fig:LB_2d_varyE}(a)],
an energy that lies just above the 
PES saddle-point energy that is
required to enable isomerization,
the LB distribution reveals a pronounced ``$\times$'' pattern that marks the primary bottleneck 
that separates the
left region that is primarily associated with 
{\LiCN} isomer
from the right region related to the
(more stable)
{\LiNC}
isomer.
As discussed before,
the large values in the LB correspond to the
invariant manifolds that emanate from
the hyperbolic point that plays the role of the
mentioned bottleneck.
Furthermore,
the LB plot has certain blueish filaments,
which correspond to other phase-space structures 
(see discussion of Fig.~\ref{fig:LB_2d_e4000},
and Sec.~\ref{sec:results_2d_sn} below).
For larger energies,
like \(4500\)\,\(\mathrm{cm}^{-1}\),
as presented in
Fig.~\ref{fig:LB_2d_varyE}(b),
the LB is also able to adequately identify the TS
along with its manifolds,
whose shapes strongly depend on the vibrational energy.
Likewise,
notice that  in this case
more structures are visible in the plot for
the considered integration time.

\begin{figure*}[t]
    \centering
    \includegraphics[width=1.8\columnwidth]{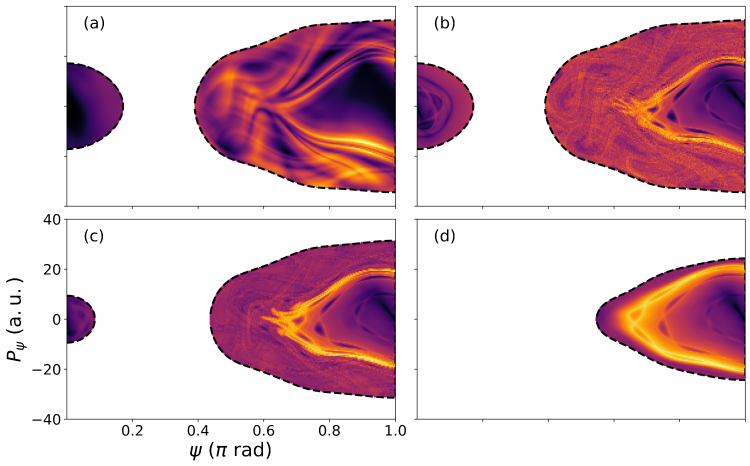}
    \caption{%
    Same as Fig.~\ref{fig:LB_2d_e4000} for
    (a) \(E=3000\,\mathrm{cm}^{-1}\), \(\tau = 10^4\) a.u.,
    (b) \(E=3000\,\mathrm{cm}^{-1}\), \(\tau = 10^5\) a.u.,
    (c) \(E=2500\,\mathrm{cm}^{-1}\), \(\tau = 10^5\) a.u.,
    (d) \(E=1500\,\mathrm{cm}^{-1}\), \(\tau = 10^5\) a.u.
    }
    \label{fig:LB_2d_time_energy}
\end{figure*}

On the contrary to the results discussed so far,
the isomerizing path keeps closed
for vibrational energies that lie under that
of the PES saddle point.
For very small vibrational energies,
the motion can only take place around the {\LiNC} well,
while other regions might also allow the motion 
around the {\LiCN} well,
but the two existing isomers of the molecular system
are fully disconnected,
being each one surrounded by a 
different hypervolumen
in phase space.
Each hypervolumen corresponds to one of the
two colored areas presented in 
Fig.~\ref{fig:LB_2d_time_energy},
which shows the LB values for 
vibrational energies
equal to
\(3000\,\mathrm{cm}^{-1}\)
[Figs.~\ref{fig:LB_2d_time_energy}(a)
and~\ref{fig:LB_2d_time_energy}(b)],
\(2500\,\mathrm{cm}^{-1}\)
[Figs.~\ref{fig:LB_2d_time_energy}(c),
and
\(1500\,\mathrm{cm}^{-1}\)
[Figs.~\ref{fig:LB_2d_time_energy}(d)].
As previously mentioned,
the accessible phase-space 
is smaller for lower vibrational energies.
In fact,
for the smallest energy considered in 
Fig.~\ref{fig:LB_2d_time_energy}(d),
only the motion of Li atom remains closer to
N than to C atoms around the stable {\LiNC} isomer
as the vibrational energy is smaller than that
of the {\LiCN} isomer 
(\(1500\,\mathrm{cm}^{-1}\)
vs \(2281\,\mathrm{cm}^{-1}\)).
As previously mentioned within the
discussion of Fig.~\ref{fig:LB_2d_e4000},
which corresponds to a vibrational energy of 4000\,cm$^{-1}$,
while the time of
\(\tau=10^4\)\,a.u.
does not show any invariant structure in 
Fig.~\ref{fig:LB_2d_time_energy}(a),
a larger computational time of
\(\tau=10^5\)\,a.u. is able to unravel chains of islands
that surround resonances,
where Li atom presents regular motion.
In fact,
the \emph{same} chains of islands,
and then resonances,
are observed.
%
Nonetheless,
the resonances experience certain
bifurcations at smaller
vibrational energies,
as the Poincaré-Birkhoff theorem dictates~\cite{LL10}.
Notice,
however,
that for 
3000\,cm$^{-1}$ [cf. Fig.~\ref{fig:LB_2d_time_energy}(b)]
and
2500\,cm$^{-1}$ [cf. Fig.~\ref{fig:LB_2d_time_energy}(c)]
the outmost chain is embedded in
the cantorus identified by the LB  in orange color.

In summary,
we have shown that the LB can be used not only
to identify the TS 
in {\LiCN}$\rightleftharpoons$LiNC isomerization,
where the reaction presents a bottleneck
limits the reaction rate
along with its corresponding manifolds,
but 
also other phase-space objects,
such as cantori and chains of islands.
In the next section,
we profound on the ability to unravel the phase space structures
that are dramatically distorted due to a saddle-node bifurcation,
which creates
an additional bottleneck of dynamical origin,
instead of energetic,
for the reaction to take place.


\subsubsection{Saddle-node bifurcation and secondary barriers}
\label{sec:results_2d_sn}

Besides the energetic
bottleneck discussed in the previous
section,
there is another one of purely dynamical origin.
Such a dynamical bottleneck is caused by
the saddle-node bifurcation that occurs at
a vibrational energy of 
$E_{\rm bif} = 3440.6$\,cm$^{-1}$,
where the gray marginally stable
PO shown in Fig.~\ref{fig:pes}
emerges out of the blue~\cite{Borondo95, Borondo96, Zembekov97, Revuelta21}.
As the vibrational energy increases,
this PO bifurcates into a stable PO (light blue) that moves to the left,
and the unstable PO (dark blue) that moves to the right.
This behavior can be observed in
Fig.~\ref{fig:pes},
where these POs at
4000\,cm$^{-1}$ (intermediate POs)
and 5000\,cm$^{-1}$ (outmost POs)
are shown.
All these POs describe a similar dynamical behavior
as they are all 1:1 resonances.

In order to study the
phase-space portrait
close to the bifurcation energy,
we illustrate in Fig.~\ref{fig:LB_SN} 
the value of the LB on the PSoS,
in the neighborhood of the region
where  the saddle-node bifurcation~\cite{Zembekov97} takes place.
Several values of the vibrational energy
are considered.
In this case,
the integration time has been increased
up to~$\tau = 2 \times 10^4$\,a.u.
in order to improve the visibility of the 
phase-space structures
responsible for
the bifurcation.

As can be seen, the LB
is able to unravel the 
invariant manifolds that determine the 
vibrational dynamics of the system.
The dynamical behavior of the molecule strongly depends on
the vibrational energy 
as it 
substantially
changes
the shape of the invariant manifolds,
which manifest as continuous
lines in Fig.~\ref{fig:LB_SN}.
Furthermore, 
their intriguing shape is responsible for the complex dynamics.

For vibrational energies below~$E_{\rm bif}$,
such as those considered on 
Figs.~\ref{fig:LB_SN}(a)-Figs.~\ref{fig:LB_SN}(c),
which respectively correspond to 
3100\,cm$^{-1}$, 3200\,cm$^{-1}$, and 3300\,cm$^{-1}$,
the LB plots are formed by several lines, i.\,e.,
invariant manifolds,
but no stability region is observed.
For an energy equal to~$E_{\rm bif}$,
some of the previous manifolds become tangent
[see Fig.~\ref{fig:LB_SN}(d)],
collapsing the stable manifolds on the unstable ones.
As a result,
a marginally stable PO appears
[cf. Fig.~\ref{fig:pes}].
This newly emerging PO manifests as
the parabolic point
shown as a gray star
on Fig.~\ref{fig:LB_SN}(d).
Due to the existence of a saddle-node bifurcation,
as the vibrational energy is only slightly increased,
the previous PO yields the two families of POs
discussed within Sec.~\ref{sec:system}:
the stable POs on the left
correspond to the
elliptic points shown as green circles
on Fig.~\ref{fig:LB_SN},
while
the unstable POs on the right
translate into the hyperbolic points
represented as a {blue crosses
on the same figure.
As the energy increases,
the size of the stability island that surrounds the elliptic points
increases.
Indeed, the island is visible 
for $E=4000$\,cm$^{-1}$ (as well as for lower energies)
on the whole PSoS
presented in Figs.~\ref{fig:LB_2d_e4000}(b) and ~\ref{fig:LB_2d_e4000}(c).
From a dynamical perspective,
the stability island corresponds to the attraction region
of the (left) stable POs which appear with the saddle-node bifurcation.
This part of the phase space is filled with an invariant torus.
Then,
any trajectory starting on an IC
located on the stable island
(or in another region where these torus is present)
will evolve on the torus for all time. 
In other words,
the stability island blocks any isomerization process,
no matter the value of the vibrational energy.

Next,
in order to assess the validity of the results of Fig.~\ref{fig:LB_SN},
we show on
Fig.~\ref{fig:FTLE_SN} the FTLE for 
the same parameters, i.\,e.,
same vibrational energy and integration time.
Both figures present a similar shape.
As a consequence,
we conclude that the LB is also able to suitably identify
this secondary bottleneck of dynamical origin.
The ability of the LB
to track emerging phase-space partitions is crucial for better understanding how vibrational dynamics evolves as energy 
surpasses
the bifurcation threshold,
in particular,
and for unveiling the reaction mechanism, 
in general.

In what follows,
we examine the performance of the LB to study a more accurate description
of the isomerization reaction based on the description of the system with its
full dimensionality.

\begin{figure*}[t]
    \centering
    \includegraphics[width=1.5\columnwidth]{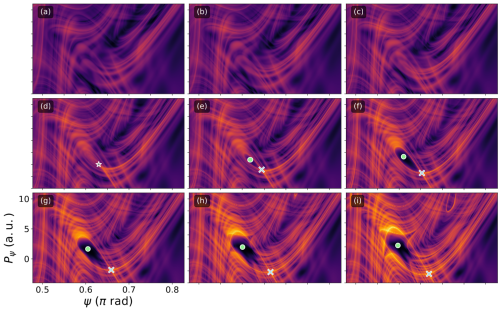}
    \caption{%
    Same as Fig.~\ref{fig:LB_2d_e4000}
    for \(\tau = 2 \times 10^4\) a.u. and
    (a) \(E=3100\,\mathrm{cm}^{-1}\),
    (b) \(E=3200\,\mathrm{cm}^{-1}\),
    (c) \(E=3300\,\mathrm{cm}^{-1}\),
    (d) \(E_{\rm bif}=3440.6\,\mathrm{cm}^{-1}\),
    (e) \(E=3500\,\mathrm{cm}^{-1}\),
    (f) \(E=3600\,\mathrm{cm}^{-1}\),
    (g) \(E=3700\,\mathrm{cm}^{-1}\),
    (h) \(E=3800\,\mathrm{cm}^{-1}\),
    and
    (i) \(E=3900\,\mathrm{cm}^{-1}\).
    The gray star,
    the green circles,
    and the blue crosses indicate
    the positions of the parabolic,
    elliptic, and hyperbolic fixed points,
    respectively, associated with
    the periodic orbits shown in Fig.~\ref{fig:pes}.
    }
    \label{fig:LB_SN}
\end{figure*}

\begin{figure*}[t]
    \centering
    \includegraphics[width=1.5\columnwidth]{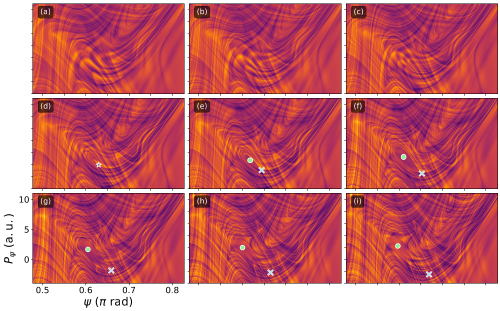}
    \caption{%
    Same as Fig.~\ref{fig:LB_SN}
    for the finite-time Lyapunov exponents.
    }
    \label{fig:FTLE_SN}
\end{figure*}

\subsection{Model with three degrees of freedom}
\label{sec:results_3d}
As previously mentioned,
the strong character of the triple CN bound effectively allows one to decouple this mode
from the rest of the vibrational dynamics,
and describe the system using the 2-DoF Hamiltonian provided by Ec.~\eqref{eq:H2dof}.
This approximation is,
however,
not possible
when all the chemical bonds lie on equal footing.
Thus,
as a starting point for the application of the LB
to identify
TSs in more complex reactions with several DoF,
we complete our study of {\LiCN}$\rightleftharpoons${\LiNC}
using the
3-DoF Hamiltonian provided by Ec.~\eqref{eq:H3dof}.
Recall that,
though the system has only one DoF more,
the complexity of the systems with 3 DoF is substantially larger
than that of systems with 2 DoF.
First,
the phase space is six- instead of four-dimensional,
and then the energy transfer between the different modes can
become much more involved~\cite{Uzer91, May11}.
Second,
the intersection of multiple resonance surfaces in action space 
might open the route for  Arnold diffusion,
a phenomenon characterized by a extraordinarily slow 
motion~\cite{Arnold64, Cincotta02}.
Third,
the PSoS,
which has an excellent performance in 2-DoF systems,
is no longer valid to
unambiguously determine if the motion 
for a given IC
is regular or chaotic~\cite{LL10, Celletti10, Cvitanovic16}.

Figure~\ref{fig:LB_3D_examples} presents LB for the full 3-DoF {\LiCN} model
under ICs
in which the kinetic energy of the \(r\) DoF
are set to zero.
Here, 
we show
the value of the LB~\eqref{eq:LB} for
uniform sets of ICs~$(\psi, P_\psi)$
on the PSoS,
each associated with a six-dimensional vector
~$\mathbf{z}_0 = (R_0, \theta_0, r_0, P_{R,0}, P_{\theta,0}, P_{r,0})$,
whose components are given by Eq.~\eqref{eq:PSoS}
once the total vibrational energy,
the initial kinetic energy in the~$r$-DoF
$\Tcn =P_r^2(0)/(2 \mu_2)$,
and the value of~$r_0$
are known.
In this case, 
the total energy is varied from \(3000\,\mathrm{cm}^{-1}\) to \(4500\,\mathrm{cm}^{-1}\),
while~$r_0 = r_e$, $\Tcn=0$,  and~\(\tau=10^4\) a.u.
First, notice the
clear similarity between the 3-DoF results and their corresponding 2-DoF counterparts
in the ``$\Tcn=0$''-case considered.
These results in a strong indication of the effective decoupling between
the $r$ mode and the rest of the DoFs.
In particular, 
the results for 
$E = 3500\,\mathrm{cm^{-1}}$ shown in 
Fig.~\ref{fig:LB_3D_examples}(b)
closely resemble those presented in 
Fig.~\ref{fig:LB_2d_varyE}(a).
The structures obtained for
$E = 4000\,\mathrm{cm^{-1}}$ in 
Fig.~\ref{fig:LB_3D_examples}(c)
bear a strong similarity to 
Fig.~\ref{fig:LB_2d_e4000}(b),
whereas the results for 
$E = 4500\,\mathrm{cm^{-1}}$ in 
Fig.~\ref{fig:LB_3D_examples}(d)
correspond closely to the patterns observed in 
Fig.~\ref{fig:LB_2d_varyE}(b).
In all these cases,
the vibrational energy surpasses the saddle point energy,
so the isomerization process is feasible;
in all these examples,
the ``$\times$'' structure associated with
the invariant manifolds of the
TS located at~$\psi \approx 0.292 \pi$\,rad
is also clearly visible.
For $E = 3000\,\mathrm{cm^{-1}}$,
which is a vibrational energy that 
does not permit isomerization,
the 3-DoF results of 
Fig.~\ref{fig:LB_3D_examples}(a)
resemble those of Fig.~\ref{fig:LB_2d_varyE}(a),
i.\,e., they do not unravel 
detailed structures such as those shown in 
Fig.~\ref{fig:LB_2d_varyE}(b).
%

\begin{figure*}[t]
    \centering
    \includegraphics[width=1.5\columnwidth]{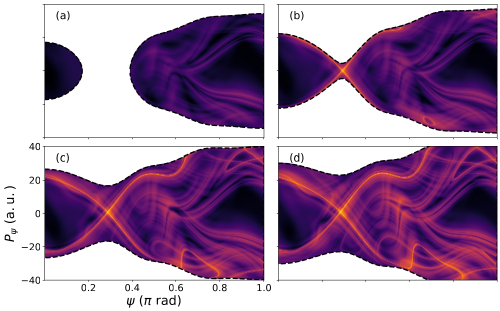}
    \caption{%
    Value of the Lagrangian betweenness~\eqref{eq:LB}
   (in logarithmic scale)
    on the Poincaré surface of section~\eqref{eq:PSoS} 
    for the 3-degrees-of-freedom Hamiltonian~\eqref{eq:H3dof}
    with no initial kinetic energy in the \(r\) coordinate (\(\Tcn=P_r^2(0)/(2\mu_2)=0\)),
    an integration time of \(\tau=10^4\)\,a.u., and a
    total vibrational energy of
    (a) \(E=3000\,\mathrm{cm}^{-1}\),
    (b) \(E=3500\,\mathrm{cm}^{-1}\),
    (c) \(E=4000\,\mathrm{cm}^{-1}\),
    (d) \(E=4500\,\mathrm{cm}^{-1}\).
}
    \label{fig:LB_3D_examples}
\end{figure*}

Figure~\ref{fig:LB_3D_various} explores a complementary scenario by fixing the total
vibrational
energy at \(4000\,\mathrm{cm}^{-1}\),
and varying~$\Tcn$.
As can be inferred from the comparison with the 2-DoF results,
the 3-DoF results resemble the 2-DoF results for energies equal to
\(4000 \,\mathrm{cm}^{-1} - \Tcn
\).
In this case,
only the 3-DoF results for $\Tcn = 500$\,cm$^{-1}$
enable isomerization,
mimicking
the 2-DoF pattern presented in
Fig.~\ref{fig:LB_2d_varyE}(a).
In this case,
the existence of a noticeable fraction of the total vibrational energy
on the $r$ DoF does not prohibit the LB
to adequately identify the ``$\times$'' structure associated with the
TS and its invariant manifolds.
However,
in this case the TS does not simply correspond to a PO
[cf. Fig.~\ref{fig:pes}] but to a
NHIM~\cite{Wiggins01, Uzer02, Waalkens04}, 
which is a more complex geometrical object of higher dimension.
Furthermore,
they are also extremely similar to the results for the 3-DoF
for~$\Tcn = 0$
[see Fig.~\ref{fig:LB_3D_examples}(b)],
which again indicates a weak coupling between the CN DoF and the~$R$ and~$\vartheta$ coordinates.

As  $\Tcn$
increases,
the effective excitation in the CN coordinate grows,
which reduces the energetically accessible area on the PSoS.
For a total vibrational energy of
E=1000\,\(\mathrm{cm}^{-1}\) [Fig. \ref{fig:LB_3D_various}(b)],
the energetically accessible region of the PSoS is formed by 
two disconnected regions,
while for the vibrational energies of
E=2000\,\(\mathrm{cm}^{-1}\) [Fig. \ref{fig:LB_3D_various}(c)]
and
E=3000\,\(\mathrm{cm}^{-1}\) [Fig. \ref{fig:LB_3D_various}(d)],
the ICs
can only be taken around the {\LiNC} isomer.
Some filaments can be observed in the plots
for moderate values in the initial kinetic energy
$\Tcn$
[Figs. \ref{fig:LB_3D_various}(b) and \ref{fig:LB_3D_various}(c)],
while the presence of a chain of islands
can be observed for the largest value [Fig.~\ref{fig:LB_3D_various}(d)].
As in the 2-DoF case,
the LB computed for longer integration times
would be able to identify other phase space structure,
such as invariant tori.
This fact can be inferred from visual inspection of Fig.~\ref{fig:LB_3D_various_t1d5},
which shows the LB for an integration time of~$\tau = 10^5$\,a.u.\,.
In this case,
the integration time is too long for
$E=4000$\,cm$^{-1}$,
and the TS is not unveiled.
Contrarily,
other phase-space structures are identified.
In particular,
a chain of large islands that surrounds the 2:2 resonance on the {\LiNC} side,
and other chains of islands on the {\LiCN} side are visible.
Similarly,
the small stability islands discussed in Sec.~\ref{sec:results_2d_sn}
is clearly identified.
Remarkably,
other islands of regularity appear embedded within the
chaotic region, 
which are absent in the 2-DoF model results shown in
Fig.~\ref{fig:LB_2d_e4000}(c).

\begin{figure*}[t]
    \centering
    \includegraphics[width=1.5\columnwidth]{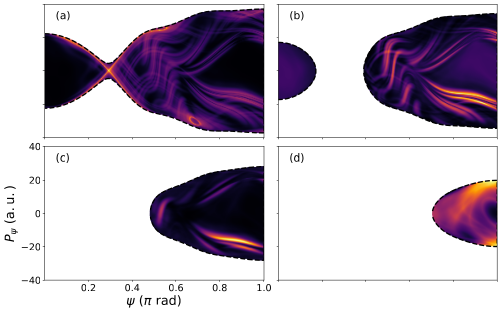}
    \caption{%
    Same as Fig.~\ref{fig:LB_3D_examples} for
    \(E=4000\,\mathrm{cm}^{-1}\), and
    (a) \(\Tcn = 500\,\mathrm{cm}^{-1}\),
    (b) \(1000\,\mathrm{cm}^{-1}\),
    (c) \(2000\,\mathrm{cm}^{-1}\),
    (d) \(3000\,\mathrm{cm}^{-1}\).
}
    \label{fig:LB_3D_various}
\end{figure*}

As in the previous instances,
when increasing the value of $\Tcn$,
the accessible area of the PSoS
is reduced.
As in the 2-DoF model,
the motion around the stable isomers is regular
in all cases.
Notice, however,
that while in the 2-DoF model
some invariant tori could be visualize around the {\LiCN} isomer
for $E=3000$\,cm$^{-1}$
[cf. Fig.~\ref{fig:LB_2d_time_energy}(a)],
that is not the case within the 3-DoF description
[cf. Fig.~\ref{fig:LB_3D_various}(b)].
In this latter case,
the invariant tori surrounding the {\LiNC}
isomer are, nonetheless,
visible as they present almost constant LB values.
As expected,
they are surrounded by invariant manifolds of unstable POs,
which appear
as the Poincaré-Birkhoff theorem predicts~\cite{LL10}.
Larger values in the $\Tcn$ energy,
further reduce the accessible 
area on the PSoS
as well as the regions of chaotic motion.
Notice in 
Fig.~\ref{fig:LB_3D_various}(c)
the orange band that lies close to the energy boundary,
which seems to have a very different dynamical behavior
compared to those points that lie in the inner part of the 
figure,
whose motion is regular.
In this case,
several chains of islands are clearly visible.
Finally,
for~$\Tcn = 3000$\,cm$^{-1}$
Fig.~\ref{fig:LB_3D_various}(d)
shows a 1:4 resonance,
being the motion for all the ICs essentially regular
due to the effective decoupling between the $r$ mode
and the rest of the DoFs in the molecule.

\begin{figure*}[t]
    \centering
    \includegraphics[width=1.5\columnwidth]{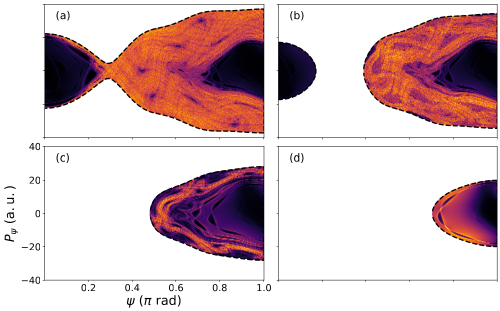}
    \caption{%
    Same as Fig.~\ref{fig:LB_3D_various} for $\tau=10^5$\,a.u.
}
    \label{fig:LB_3D_various_t1d5}
\end{figure*}

Overall,
the 3-DoF results confirm that LB is not only able to identify the TS for the
2-DoF model of the isomerization reaction but also 
in this higher dimensional case,
as well as other phase-structures which are responsible for the 
occurrence of regular motion.

\section{Conclusions and Outlook}
\label{sec:conclusions}

In this work, we have explored the use of Lagrangian betweenness to identify transition states and dynamical bottlenecks in the
{\LiNC}\(\leftrightharpoons\){\LiCN}
isomerization reaction. 
Lagrangian betweenness is motivated by betweenness centrality in network theory, but applied here to continuous phase-space flows by measuring how many trajectories pass through a given region over a finite time.
We have showed that the Lagrangian betweenness is suitable for reduced
(2-degrees-of-freedom) and a full (3-degrees-of-freedom)
realistic Hamiltonian models of {\LiCN} isomerizing system.

First,
our numerical studies confirm that Lagrangian betweenness is particularly effective at highlighting the
hyperbolic invariant manifolds controlling the reaction, 
which emerge from the transition state
located at the top of the energetic barrier
that separates the two stable isomers of the system.
The transition state determines a strictly recrossing free dividing surface
that is crossed once and only once for all reactive trajectories.
In the case of the 2-degrees-of-freedom model,
the transition state corresponds to an unstable periodic orbit~\cite{Pechukas73, Pollak78, Pechukas79},
while for the 3-degrees-of-freedom model,
and in higher dimension,
the transition state corresponds to a NHIM~\cite{Wiggins01, Uzer02, Waalkens04}.
Likewise,
the good performance of the 
2-degrees-of-freedom model has been assessed
by comparison
of its results with those of the 
3-degrees-of-freedom model,
which are very similar if the initial kinetic energy in the $r$ mode
is subtracted.

Second,
we have shown that the Lagrangian betweenness
is also able to unravel another reactive bottleneck of dynamical origin that inhibits
isomerization.
Such a bottleneck is caused by a saddle-node bifurcation which creates a
stability region within the chaotic sea~\cite{Borondo95, Borondo96, Zembekov97, Revuelta21}.
Consequently,
if the system is prepared on that region of the the phase-space,
no reaction is possible,
as any trajectory would remain on an invariant torus for all time.
We have studied the dependence on the vibrational energy of the phase-space structures
involved in the bifurcation,
observing that the stability island can be easily identified as
a region with an almost constant value of the Lagrangian betweenness.
Moreover,
this region is surrounded by other invariant manifolds,
which form the homoclinic tangle that Poincaré foresaw~\cite{Arnold06}.

Third,
we have studied the effect of the integration time on the performance of
the Lagrangian betweenness.
As in the case of other well-established chaos indicators
which have been taken as reference,
such as 
the finite-time Lyapunov exponents~\cite{Haller01, Haller02}
or the Lagrangian descriptors~\cite{Madrid09, Lopesino15, Revuelta19, Revuelta21},
when considering a \emph{too} short time
for a uniform set of initial conditions,
the representation of the corresponding 
Lagrangian betweenness yields a slow-varying function
where no pattern is recognized,
while \emph{excessively} long times
includes too much dynamical information
rendering a \emph{pointillist}-like plot,
where the transition state cannot be correctly identified.
The proper integration time is related to the inverse of the
stability exponents of the transition state:
if the integration time is too short,
the Lagrangian betweenness  does not have enough time to
account for the hyperbolic behavior of the transition state,
while too long times 
account for the local information of the invariant manifolds of
interest, as well as of other phase-space structures.
Remarkably,
long integration times can be used to identify other 
phase-space structures such as
the invariant tori that, for the case study,
surround the two stable isomers of {\LiCN} molecular system,
as assessed by comparison with the
results obtained with a characteristic Poincaré surface of section~\cite{LL10, Celletti10, Cvitanovic16}
defined along the minimum energy path of the system.

To conclude,
let us remark that the studied system strongly relies on the
validity of Born-Oppenheimer approximation,
where the atomic nuclei sojourn the 
potential energy surface created by the electrons.
However,
it would be also interesting to test the validity of the
Lagrangian betweenness where this adiabatic approximation
fails~\cite{Baer06},
due to the existence of
conical intersections~\cite{Yarkony96, Domcke04},
nonadiabatic transitions~\cite{Fiebig01, Bokang23},
or
strong fields~\cite{Lutz16, Liu19},
among other phenomena.
Notice, also,
that the current definition of the Lagrangian betweenness is purely
classical,
but certain reactions, such as those involving light particles,
require a quantum description.
Other effects,
such as couplings with the environment or 
the inclusion of external fields will be considered in the future.

\section*{Acknowledgments}

This work has been partially supported by the Grant PID2021-122711NB-C21
funded by MCIN/AEI/10.13039/501100011033.
The authors acknowledge computing resources at the Magerit Supercomputer of the
Universidad Polit\'{e}cnica de Madrid.

\section*{Declaration of generative AI and AI-assisted technologies in the writing process}
Generative AI tools were used to improve the clarity and readability of the manuscript.
All content generated using these tools was critically reviewed, verified, and edited by the authors,
who accept full responsibility for the final version of the article.


\bibliography{lb_licn}

\end{document}